\documentstyle[preprint,aps]{revtex}
\begin{document}
\author{S.F. Burlatsky\cite{Auth}\dag}
\author{J.G. Berberian\ddag}
\author{J. Shore\ddag}
\author{W. P. Reinhardt}
\title{Spreading of a Macroscopic Lattice Gas }
\address{Department of Chemistry, 351700,\\
University of Washington,\\
Seattle, WA 98195-1700,\\
USA \\
Department of Physics, Saint Joseph`s University, 5600 City Avenue\\
Philadelphia, PA\\
19131-1395, USA \\
\dag e-mail: sfburlat@ringa.chem.washington.edu; sfburlat@sjuphil.sju.edu.}
\maketitle

\begin{abstract}
We present a simple mechanical model for dynamic wetting phenomena. Metallic
balls
spread along a periodically corrugated surface simulating molecules of
liquid advancing along a solid substrate. A vertical stack of balls mimics a
liquid droplet. Stochastic motion of the balls, driven by mechanical
vibration of the corrugated surface, induces diffusional motion. Simple
theoretical estimates are introduced and agree with the results of the
analog experiments, with numerical simulation, and with experimental data
for microscopic spreading dynamics.
\end{abstract}

\date{}
\draft

\section{Introduction}

Because of its practical applications in areas such as coating, lubrication,
adhesion, etc., the old field of wetting phenomena has recently attracted
renewed interest \cite
{CFVH,Gran,KaskiAbraham,Kaski,AbrahamKaski,GennesRevMod,Cazabatrev,JoannyGennes}
Micro droplets which spontaneously spread along a solid surface have a time
dependent shape which results from the balance between liquid - solid
interactions and friction processes\cite{GennesRevMod}. Thus wetting
mechanisms are intimately connected with friction on a molecular level \cite
{GennesRevMod,StickSlip,ThompsonGrest} which is also important in the
understanding of such practical problems as friction between two solid
surfaces separated by a thin liquid layer\cite{Gran} and dynamics of long
polymer chains in random media\cite{SFSci}.

A salient feature of macroscopic spreading is that it is often preceded by a
microscopically thin film - precursor. The precursor film thickness may
vary from molecular size (one or sometimes several monolayers) to a few
hundreds of angstroms. \cite{har}. For nonvolatile liquids, well below their
critical temperature, thickness profiles with distinct successive molecular
layers (terraces)  have been observed \cite{wetting,VFCH}. Ellipsometric
measurements,
carried out on different substrates and also for various kinds of simple
liquids, as well as polymeric and surfactant melts, have reached a
surprising conclusion: the linear size $R$ of the precursor obeys a
universal law \cite{wetting,tib}
\begin{equation}
R\propto \sqrt{t},  \label{sqrt}
\end{equation}
$t$ being the time. The same law holds also for capillary rise, in which a
vertical wall is put into a contact with a bath of liquid. Here a film of
microscopic thickness grows from the macroscopic liquid meniscus and creeps
upwards along the wall. In this case, the height of the film obeys the $%
\sqrt{t}$-law within an extended time domain \cite{hesd1,hesd2}, until it
gets truncated,. at very high altitudes, by gravity. A diffusion-like
coefficient $D_1$ can be formally defined as a prefactor in the Eq.~(\ref
{sqrt}), which is found to scale as the inverse of the bulk viscosity. Such
a formal fitting does not, however, immediately imply an understanding or
even a model of the microscopic dynamics responsible for the simple power law%
\cite{Cazabatrev}.

Theoretical understanding of precursor dynamics has followed from two major
conceptualizations. The hydrodynamic approach (HA) \cite{CazabatdeGennes},
and the Solid on Solid Model (SOSM) \cite{Abragam} which is based on
Langevin equations for layers in the drop. Both approaches qualitatively
describe the formation of layered structures. HA correctly describes long
term kinetics of terraced spreading in $2D$ systems with cylindrical
symmetry, however the SOSM predicts $R\propto t$ rather then$\sqrt{t}.$ In
both models the layers are considered as being incompressible continua
neither assumes nor implies a microscopic model of the dynamics. An analogy
with the
analysis of macroscopically thin layers \cite{JoannyGennes}, and Ising -
like models \cite{KaskiAbraham}, as well as the diffusion like structure of
Eq. (1), suggests that diffusion inside the precursor layer plays an
essential role for spreading dynamics. However consideration of the motion
of precursor edge (PE) as a simple biased random walk is incorrect\cite
{JoannyGennes}. In the presence of an external force (capillary force which
pulls the precursor out of the drop\cite{Cazabatrev}), the mean displacement
of a biased random walker (at the PE) is proportional to time in violation
of Eq. (1), as is the prediction for the dynamics of the first layer in the
SOSM \cite{Abragam}). In the absence of external forces the square root of
mean squared displacement would exhibit the behavior of eq. (\ref{sqrt}).
However, the mean PE displacement would be zero contrary to wetting
experiments which indicate \cite{Cazabatrev} a continuous directed
displacement of the PE with relatively small fluctuations. This differs
strongly from the fluctuation induced motion of a non-biased random walker,
where the fluctuations are of the same order as the typical displacement.

In \cite{DS,BOMR1} it was shown that the mean displacement of a random
walker which is biased by a uniform external force and additionally
experiences excluded volume repulsion exerted by an ensemble of other, non
biased, diffusing hard core particles grows in proportion to $\sqrt{t}$,
instead of the linear in time growth expected for similar systems without
the hard core repulsion. We also have shown in this model that excluded
volume effects imply an effective {\em frictional} force imposed on the
motion of an individual particle in the hydrodynamics description.

The goal of the present work is to investigate a simple macroscopic
spreading process with excluded volume. We consider the spreading of
metallic balls, with and without magnetic interactions between the balls,
and find that the spreading of balls in 1D belongs to the same universality
class as spreading of microscopic particles of liquids, i.e. is governed by
eq. (\ref{sqrt}). We present experimental results for a the spreading of
macroscopic balls, which emerge from a reservoir (vertical stack) , and
spread onto a horizontal vibrating rack, see Fig. \ref{eseutup}. The surface
of the rack is uniformly corrugated to prevent rolling or ballistic motion.
of the balls. Constraints enforce zero or unit occupancy of corrugated
sites, creating a lattice system. Weak random driving leads to {\em lattice
gas} type behavior of balls - random jumps of length +/-1 constrained by the
excluded volume effect. We find that
this analog model yields the $\sqrt{t}$-law for the total number of balls which
emerge from the reservoir, for the average displacement of balls from the
reservoir origin, and for the displacement of the rightmost ball. We
also present analytical estimates and numerical simulations, which are in
good agreement with experimental data. The results suggest that, as far as
time dependence is concerned, the $\sqrt{t}$-law is essentially independent
of the nature of the interactions between the substrate and the spreading
substance, long range interactions between the particles themselves, as well
as of the geometry and of the size of spreading particles. The prefactors in
the square root law, of course, do depend on the system's parameters and all
details of
the microscopic interactions.

The organization of the paper is as follows: In the Sections \ref{est} and
\ref{numpr}we describe the experimental set-up and numerical algorithm, in
the Section \ref{anest} we present a simple analytical results for our
model: basic equations - in the Section \ref{be}, results for the flux of
balls from the reservoir and averaged displacement of spreading balls in the
Section \ref{flavd}, results for the displacement of the rightmost ball,
that determines the size of the spreading layer - in the Section \ref{rmb},
and separate results for the steady regime which starts after the firs ball
falls out of the rack - in the Section \ref{stead}. In the Section
\ref{results} we present experimental and numerical results, which are briefly
summarized in the Section - \ref{summary}.

\section{Experimental set-up and numerical procedure}

\subsection{Experiment}

\label{est}A schematic diagram of the experimental set-up appears in Fig.\
\ref{eseutup}. A corrugated (notched) horizontal rack is confined inside a
rectangular tube which prevents balls from passing one another or jumping
off the track. A vertical stack is placed at the left end of the rack (at
the origin) and metallic balls (balls) are fed through this stack
maintaining unit concentration of `particles' at the origin. The balls are
allowed to move to the right of the origin only. The entire system is driven
with motors placed on each end of the rack. The flywheel on each motor is
eccentric to provide ''chaotic'' oscillations which are experimentally shown
(see below) to give rise to diffusional motion of the balls. A particular
number of balls, $n$ $(n=4$ or $8)$, is placed in the vertical stack and
then the oscillations are started. The time for the balls to leave the stack
is measured and recorded along with the displacement of the horizontal
balls. After this, $n$ more balls are added to the stack and the procedure
is repeated with the first $n$ balls left at their respective place on the
horizontal rack. The process is repeated until the rightmost ball reaches
the right end of the rack. Results presented are averaged with respect to
four independent trials. In a separate set of experiments the number of
balls which emerge from the vertical stack as a function of time after the
rightmost ball reached the right end of the rack and escapes from the rack
was measured. This is the spreading rate for a ''full'' horizontal rack. Two
kinds of balls we used: magnetic and non magnetic.

The experiment was designed to mimic the spreading of liquids: the vertical
stack imitates a liquid drop, which acts as a reservoir providing particles
for the precursor and keeps constant concentration at the drop - precursor
boundary, the horizontal rack mimics a solid surface. Since the gravitation
energy of balls on the rack is proportional to the height, the shape of the
surface of the rack emulates the profile of the potential of liquid -
substrate interactions with local potential minima, which prevent particles
from long range (on length scales larger then the mean distance between two
neighboring particles) ballistic or rolling motion. Oscillations of the
horizontal rack mimic thermal excitations of the surface of a solid
substrate and generate random jumps of balls along the rack. The number $n$
determines the gravitation induced pressure in the vertical stack. Two sets
of experiments - with $n=4$ and $n=8$ were carried out in order to make sure
that in given range of $n$'s values the pressure is not important, see
section \ref{results}. The total number of balls on the horizontal stack, $%
M(t)$, is an analog of the precursor mass and the displacement of the right
most ball is an analog of the precursor radius, $R(t)$. The magnetic mutual
ball to ball attraction serves to mimic the particle - particle attraction
in a precursor film and in a liquid drop.

To test the assumption that the eccentric oscillations generate diffusive
motion of a single ball, Fig.\ \ref{tsetup} shows the mean squared
displacement and the squared mean displacement of a single ball initially
placed on the horizontal rack half way between the origin and right-most end.
The observed linear dependence of the mean squared displacement on time
agrees with well known result of conventional single particle random walk
theory. The observed RMS dispersion from the mean is also in approximate
agrement with the prediction of simple estimates presented in the Appendix.
However, the absolute values of dispersion is approximately $1.5$ times
greater then the theoretical value. These experimental result indicates that
the jumps of a single ball are, to a good approximation, random and
independent events caused by the vibrational driving of the system.

\subsection{Numerical simulations}

\label{numpr}In numerical simulations, we modeled random jumps of particles
on a $1D$ lattice with unit steps. For each time, each particle except the
particle in the site number $0$ chooses randomly a direction: to the right
or to the left with equal probabilities. If the corresponding neighboring
site is occupied, the move is rejected and the particle stays at the
original position; if the site is vacant, the particle jumps to it. The
origin (site 0) is always occupied; a new particles is automatically added
to the site number 1 when it becomes vacant. To simulate an analog of
magnetic interaction in the simplest manner, the jump rates of particles
which had a nearest neighbor from one side to the vacant site on the other
side was reduced by the factor $2(1-p)$ where $p$ determines the ''strength
of the interaction'', $p=0.5$ corresponds to the absence of the interactions
and $p=1$ corresponds to infinitely strong attraction. At an initial time,
two particles are placed into the system: the first particle - at $i=0$ and
the second particle - at $i=1$. The mean displacement of balls was measured
as was the number of balls as function of time up to 100 active balls on the
rack. Results were averaged with respect to $40$ independent trials for each
$p=0.5,0.6,0.7,0.8.$

\section{Simple analytical estimates}

\label{anest}

\subsection{Diffusion equation and boundary conditions}

\label{be}In order to obtain a simple analytical estimate for the dynamics
of the processes we neglect mutual magnetic attraction (this is the same as
neglecting  surface tension for microscopic spreading phenomena), and
decouple many particle probabilities. This leads to following equations for
mean concentrations in $1D$ lattice with unit step (horizontal rack):

\begin{equation}  \label{1}
\begin{array}{c}
\frac{\partial C_v(i,t)\ }{\partial t}=\frac \omega 2\{C_v(i+1,t)[1-C_v(i,t)]
\\
+C_v(i-1,t)[1-C_v(i,t)]-C_v(i,t)[1-C_v(i-1,t)] \\
-C_v(i,t)[1-C_v(i+1,t)]\}
\end{array}
,
\end{equation}
where $\omega $ is frequency of jumps, which is connected to the diffusion
coefficient, $D=\frac{\omega l^2}2,$ $l$ is the jump length, $C_v(i,t)$ -
concentration of vacancies (empty slots on the horizontal rack) at the slot
number $i$ at time $t,$ $C_v(i,t)=1-C_b(i,t),\,C_b(i,t)$ is the probability
to find a ball (ball) at slot number $i$ at time $t$. The first term in the
right hand side of the Eq. (\ref{1}) determines the rate of jumps of
vacancies from the site number $i+1$ to the site number $i$. It is
proportional to the concentration of vacancies at the site number $i+1$
multiplied by the concentrations of particles at the site number $i$ since
only vacancy - particle exchanges are allowed. The other terms determine the
rate of jumps $i-1\rightarrow i,$ and $i\rightarrow i-1$ and $\,i\rightarrow
i-1$ in a similar way.

In the approximate Eq. (\ref{1}), nonlinear terms cancel since the forbidden
particle - particle and vacancy - vacancy exchanges do not alter the local
mean concentrations. When $i$ and $t$ are large, $i$ $>>1$ and $\omega t>>1$%
, it leads to the diffusion equation:
\begin{equation}  \label{diffeq}
\frac{\partial C_v(i,t)}{\partial t}=\frac \omega 2\Delta C_v(i,t),
\end{equation}
where $\Delta $ is Laplace operator with respect to the variable $i$. At $%
t=0 $ and $\ i>0$, there is no ''liquid'' on the solid surface (no balls on
the horizontal rack). Therefore, the initial concentration of vacancies is
equal to unity:
\begin{equation}  \label{ic}
\left. C_v(i,t)\right| _{t=0}=1
\end{equation}
- all sites are empty. When $i$ is large the concentration of vacancies is
equal to unity because the balls have not had enough time to reach this
area,
\begin{equation}  \label{infbc}
\lim _{i\rightarrow \infty }C_v(i,t)=1.
\end{equation}
There are no vacancies at the boundary $i=0$, because the reservoir of balls
places a ball in the vacancy instantly,
\begin{equation}  \label{zerobc}
\left. C_v(i,t)\right| _{i=0}=0.
\end{equation}
Note that in spite of the fact that the Eq.~(\ref{diffeq}) apparently does
not reflect excluded volume interaction, the boundary condition Eq.~(\ref{ic}%
) states that the ball from the vertical stack can move down to the
horizontal rack if and only if the vacancy comes to the point $i=0$. When
the ball falls down from the stack it eliminates the vacancy at $t=0$. The
relaxation time for the vacancy concentration in the horizontal stack is
determined by the diffusion of a vacancy through the array of balls in the
horizontal stack - from the right end to the left end. Since the number of
balls and the length of the array grow, the relaxation time increases which
slow down the dynamics and leads to the dependence presented by Eq.~(\ref
{sqrt}) instead of linear growth.

\subsection{Flux from reservoir, number of balls, and averaged displacement}

\label{flavd}The solution of the Eq.~(\ref{diffeq}) in $1D$ with the
boundary conditions eq.~(\ref{ic}) and eq.~(\ref{infbc})
\begin{equation}
C_\nu (i,t)=1-erfc(\frac i{2\,\sqrt{Dt}})  \label{sol}
\end{equation}
leads to the following result for the flux, $P(t)$, at $i=0$
\begin{equation}
P(t)=\left. \frac \omega 2\frac{\partial C_v(i,t)}{\partial i}\right| _{i=0}=%
\sqrt{\frac \omega {2\pi t}}.  \label{1dflux}
\end{equation}
The number of particles on the rack is equals to
\begin{equation}
M(t)=\int\limits_0^tP(\tau )d\tau =\sqrt{\frac{2\omega t}\pi }.
\label{Npart}
\end{equation}
Writing down the equation similar to Eq.~(\ref{1}) for particles
concentration, multiplying both sides of this equation by $i$ and
integrating from $0$ to infinity, and taking into account the boundary
conditions Eq.~(\ref{zerobc}) and Eq.~(\ref{infbc}), we obtain, for the mean
total displacement of all balls in the horizontal rack, $K(t)=\int\limits_0^%
\infty iC_b(i,t)di,$ measured in the jump length units $l$%
\begin{equation}
\frac{dK(t)}{dt}=\frac 12\omega .  \label{dkt}
\end{equation}
Integrating Eq.~(\ref{dkt}) and taking into account the normalization
\[
M(t)=\int\limits_0^\infty C_b(i,t)di,
\]
we obtain the averaged over all balls displacement for the balls in the
horizontal rack, $r(t)$%
\begin{equation}
r(t)\equiv \frac{K(t)}{M(t)}=\frac 14\sqrt{2\pi \omega t}.  \label{avdispl}
\end{equation}

The straightforward extension of the results, Eq.~(\ref{Npart}) and Eq.~(\ref
{avdispl}) for the corresponding $2D$ system in large $t$ limit leads to
\begin{equation}  \label{2dflux}
\left. P(t)\right| _{2D}=t\left. \frac{2\pi a\omega }2\frac{\partial C_v(i,t)%
}{\partial r}\right| _{i=0}\propto \frac{\pi \omega }{2\ln (\frac{Dt}{a^2})}.
\end{equation}
and
\begin{equation}  \label{2ddispl}
\left. M(t)\right| _{2D}\propto \left. r(t)\right| _{2D}\propto \frac{\pi
\omega t}{2\ln (\frac{Dt}{a^2})}.
\end{equation}

\subsection{Displacement of the rightmost ball}

\label{rmb}One can show \cite{BOMCR} that apart from logarithmic
corrections, which can occur in long time regime of spreading, the
displacement of the rightmost ball on the horizontal rack (the length of the
precursor- $R(t)$) is proportional to the average displacement, $R(t)\propto
r(t)$, which leads to the Eq. (\ref{sqrt}). Here we present a simple
estimate for the $R(t)$. The distribution of the displacement of the right
most particle, for the ensemble of spreading hard core particles can be
bounded by the maximum displacement of independent diffusing particles. The
later displacement has the following distribution
\begin{equation}
P(R)=c(R,t)e^{-\frac 1a\int_0^R\!\ln (1-c(r,t))dr},  \label{distrR}
\end{equation}
where $c(R,t)=1-C_\nu (i,t),$ $C_\nu (i,t)$ being determined from the Eq.~(%
\ref{sol}). The first multiplier in the right hand side of the Eq.~(\ref
{avdispl}) is the probability to have a particular displacement $R$ for a
ball, the second term is a limiting continuum form for a probability to have
smaller displacement for other balls. Averaging $R$ with respect to the
distribution Eq.~(\ref{distrR}) by means of steepest decent method in large $%
t$ - limit we obtain
\[
<R>=\sqrt{2Dt}\sqrt{\ln (\frac{16\,(Dt)^3}{\pi a^2})}.
\]

\subsection{Steady regime}

\label{stead}The regime of spreading changes after the first ball falls off
of the horizontal rack. The length of the array of balls does not grow any
more and the concentration of balls is determined by the steady state
solution of the Eq.~(\ref{diffeq}) with a new boundary condition at the
right side of the horizontal rack
\begin{equation}
\left. C_v(i,t)\right| _{i=L}=1,  \label{rightbc}
\end{equation}
where $L$ is the total number of slots on the horizontal rack. This boundary
condition states that there are no balls out of the rack with the
coordinates $i>K$. The solution of the steady state Eq.~(\ref{diffeq}) leads
to the constant flux of vacancies to the origin, which in turn leads to
\begin{equation}
M_1(t)=\frac \omega {2L}(t-t_0),  \label{line}
\end{equation}
where $M_1$is the number of balls which fall out of the rack and $t_0$ is
the time when the first ball falls.

\section{Experimental and numerical results}

\label{results}The Fig.\ \ref{ballonsqrt} and Fig.\ \ref{displsqrt} show the
experimental results for the number of balls in the horizontal rack and the
averaged displacement of the right most ball versus square root of time for
magnetic and non magnetic balls for $n=4,8$ and $t<t_0$,

Fig.\ \ref{extter}
shows the time for a given number of balls to enter the horizontal
rack from the stack .  In this case the experiment was not interrupted at time
$t=t_0$.

The Fig.\ \ref
{msqsqm} presents the dependencies of the mean squared displacement along
with the squared mean displacement of the rightmost ball in the horizontal
rack. The figures Fig.~\ref{simnpart}, Fig.~\ref{simdispl}, and Fig.~\ref
{simmaxdispl} show the dependencies of the mean number of balls in the
horizontal rack, mean displacement of balls on the horizontal
rack, and the mean displacement of the right most ball versus the square root
of time obtained in the numerical simulation for different values of the
''interaction parameter'' $p$.

The pronounced straight lines which are presented in Fig.\ \ref{ballonsqrt},
Fig.\ \ref{displsqrt}, Fig.\ \ref{msqsqm}, Fig.\ \ref{simnpart}, and Fig.\
\ref{simdispl} are in good qualitative agreement with Eq.~(\ref{avdispl}),
Eq.~(\ref{Npart}) and Eq.~(\ref{sqrt}). However, the fluctuations in the
dependence of the mean displacement on square root of time are larger then
fluctuations for the dependence of the total number of balls. The results
for $n=4$ and $n=8$ are essentially similar. It shows that the gravity -
induced pressure in the reservoir does not play significant role, which
corresponds to the spreading of small droplets or vertical creep with
relatively small precursor length where the gravitational forces
are not important. In a test experiment with larger number of balls, $n\geq
40$, the large gravitational force pushed all balls from the vertical stack
after beginning of vibrations, overcoming the potential barriers produced by
slots on the horizontal rack, i.e. large gravitational forces changed the
nature of spreading.

The theoretical values for the slopes  of the linear dependencies of the number
of
balls and of the mean displacement of balls in the precursor which are
determined by the Eq.~(\ref{Npart}) and Eq.~(\ref{avdispl}) for $\omega =1$:
$\sqrt{\frac 2\pi }\approx 0.798$ and $\sqrt{2\pi }/4\approx 0.6266,$ are in
excellent agreement with the corresponding values determined from the
numerical experiment : $0.804$ and $0.625$.

Magnetic interaction does not change the shape of the dependencies but
decreases the numerical prefactors and increases fluctuations. Introduction
the effective interaction in numerical simulations leads to similar effects.
The numerical results for the dependencies of the mean number of balls on
the horizontal rack and of the mean displacement versus square root of time,
presented in Fig.\ \ref{simnpart} and Fig.\ \ref{simdispl} exhibit a slowing
down when the ''interaction parameter'' increases from $0.5$ to $0.8$. The
fluctuations in the mean squared displacement also increase and crossover
period, which can be seen for small times in Fig.\ \ref{simdispl} becomes
larger.

It is also instructive to compare the results presented in Fig.\ \ref{tsetup}
for the moments of the displacement of the single ball on the horizontal
rack with the results for the moment of the displacement of the rightmost
ball of the array spreading along the rack, presented in Fig.\ \ref{msqsqm}.
For a single ball the mean squared displacement is proportional to time
while the squared mean displacement is much smaller and irregular. In the
limit of large number of trials, $K,$ it should tend to zero as $1/\sqrt{K}$
On  the contrary, the experimentally measured mean squared displacement and
squared mean displacement for the rightmost ball in the array are equal
within the error of the experiment. This shows that, in spite of apparent
scaling, the similarities of these two processes are of essentially
different physical origin and behavior. The motion of a single ball is a
fluctuation induced process with zero mean, while the spreading is a driven
diffusive process with small fluctuations.

The initial parabolic dependence presented in Fig. 5 for $t<t_0$ agrees with
the Eq.~(\ref{Npart}) while the linear dependence obtained for $t>t_0$
confirms Eq.~(\ref{line}). The numerical prefactors for these time dependencies
 are
not the subject of our present work;  thus actual values of the
vibrational frequencies and length scales are not important for our results.
However,  we can suggest some rough estimates, which produce reasonable
numerical values. The effective experimental frequency of jumps $\omega
\approx 32sec^{-1}$ was roughly estimated from the data presented in Fog. 5
for the time $t>t_0$ (the liner regime) with the approximate values for the
length of the rack $\approx 60cm$ and $l\approx 1cm$. It is in good
agreement with the values $\omega \approx 36sec^{-1}$ which was obtained by
independent estimate from the region $t<t_0$ my by means of Eq.~(\ref{Npart}%
). The slope for the experimental dependence of mean displacement versus
square root of time is also in good agreement with the theoretical estimate.

\section{Summary}

\label{summary} We have calculated analytically, simulated numerically, and
measured experimentally the mass and the size of a wetting layer for a
simple mechanical system - metallic balls, which spread along a periodically
corrugated surface, which mimics the wetting phenomena. The experimental
results and numerical data are in excellent agreement with the theoretical
predictions concerning the dependencies which determine the dynamics of
spreading of a lattice gas (balls) in our mechanical model\cite{note} .
These agree with Eq.~(\ref{sqrt}) which empirically describes the dynamics
of precursors in spreading of actual liquid drops. The mechanical model may
thus reflect the important features of this phenomena which lead to the
universal spreading law, eq. (\ref{sqrt}). Analysis of the experimental
results, simple analytical estimates and numerical simulations provides an
additional argument that the precursor kinetics is controlled by the
diffusion of vacancies from the precursor's boundary to the liquid drop \cite
{CFVH,Gran,KaskiAbraham,GennesRevMod,BOMCR}. A theoretical model which
directly explores the consequences of these ideas for a model of microscopic
wetting is presented in a subsequent paper \cite{BOMCR}, which substantially
extends the ideas of \cite{BOMR}.

\acknowledgments
The authors thank S. Granick, A.M. Cazabat, M. Robbins, M. Moreau, and G.
Oshanin for helpful discussions. This work was supported in part by ONR
Grant N00014-94-0647.

\appendix

\section{Dispersion of mean squared displacement}

\label{append}Let us consider the dispersion for the average over $K$
realizations squared displacement of a random walks containing $N$ steps.
What we measure is:

\begin{enumerate}
\item  mean squared displacement for individual random walk
\[
<R_i^2(N)>=<\left( \sum_{j=1}^Ns_i\right) ^2>=\sum_{j=1}^N<s_j^2>,
\]
where $s_j$ is the step number $j$, $R_i^2(N)$ is the averaged squared
displacement of the random walk number $i$.

\item  squared displacement for individual random walk realization
\[
R_i^2(N)=\left( \sum_{j=1}^Ns_i\right) ^2,
\]

\item  averaged with respect to all realizations squared displacement
\[
\frac{\sum\limits_{j=1}^KR_i^2(N)}K=\frac{\sum\limits_{j=1}^K\left(
\sum_{j=1}^Ns_i\right) ^2}K,
\]

\item  square root from averaged deviation of mean squared displacement from
it's averaged value
\[
\Delta =\sqrt{\frac{\sum\limits_{i=1}^K\left( R_i^2(N)-\frac{%
\sum\limits_{i=1}^KR_i^2(N)}K\right) ^2}K},
\]
Denote $R_i^2(N)=X_i$
\[
\Delta ^2=\frac{\sum\limits_{i=1}^K\left( X_i-\frac{\sum\limits_{i=1}^KX_i}K%
\right) ^2}K=\frac{\sum\limits_{i=1}^K\left( X_i\right)
^2-\sum\limits_{i=1}^K\left( \frac{\sum\limits_{i=1}^KX_i}K\right) ^2}K
\]
\end{enumerate}

All $R_i(N)$ are independent random variables with zero mean and second
moment equal to $N\sigma ^2$, the gaussian random variables $R_i^2(N)$ have
mean equal to $N\sigma ^2$, and second central moment equal to $3N^2\sigma
^4.$ For $K=30,\sqrt{30}\propto $ $5.4772,$ $1/\sqrt{30}=.18257$ and $\Delta
=\sqrt{3/\sqrt{30}}N\sigma ^2=.74008N\sigma ^2;$

\bibliographystyle{plain}
\bibliography{reinstu}

\begin{thebibliography}{10}

\bibitem{Auth}
{Auth} {\em Permanent address:\/} Institute of Chemical Physics, Russian
  Academy of Sciences, Moscow 117977, Russia.  .

\bibitem{CFVH}
J.~D. Coninck, N. Fraysse, M. Valignat, and A. Cazabat, Langmuir {\bf 9},  1906
   (1993).

\bibitem{Gran}
S. Graninck, {\em Molecular tribology of fluids} in " Fundamentals
of friction",  ed. by I.L. Singer and H.N. Pollock,
NATO ASI proceedings,  Kluwer, Dordrecht, Boston, London, (1992).

\bibitem{KaskiAbraham}
A.~L. K. K.~D. Abraham, Phys. Rev. E {\bf 51},  51  (1995).

\bibitem{Kaski}
K. Kaski, Europhys. News {\bf 26},  23  (1995).

\bibitem{AbrahamKaski}
D.~A. K.Kaski, Proceedings of Les Houches School on Dynamical Phenomena at
  Interfaces, Surfaces and Membranes  (1993).

\bibitem{GennesRevMod}
P. de~Gennes, Rev. Mod. Phys. {\bf 57},  827  (1985).

\bibitem{Cazabatrev}
A.~M. Cazabat, Contemp. Phys. {\bf 28},  347  (1989).

\bibitem{JoannyGennes}
J.~J.~P. de~Gennes, J. Phys (Paris) {\bf 47},  121  (1986).

\bibitem{StickSlip}
R. Overney {\it et~al.}, Phys. Rev. Lett. {\bf 72},  3546  (1994).

\bibitem{ThompsonGrest}
P. Tompson, G. Grest, and M. Robbins, Phys. Rev. Lett. {\bf 66 (68?)},  3448
  (1992).

\bibitem{SFSci}
S. Burlatsky and J. Deutch, Science {\bf 260},  1782  (1993).

\bibitem{har}
W.~B. Hardy, Phil. Mag. {\bf 38},  49  (1919).

\bibitem{wetting}
F. Heslot, N. Fraysse, and A. Cazabat, Nature {\bf 338},  640  (1989).

\bibitem{VFCH}
M. Valignat, N. Fraysse, A. Cazabat, and F. Heslot, Langmuir {\bf 9},  601
  (1993).

\bibitem{tib}
F. Tiberg and A.~M. Cazabat, Europhysics Letters {\bf 25},  205  (1994).

\bibitem{hesd1}
F. Heslot, A.~M. Cazabat, and N. Fraysse, J. Phys. Cond. Matter {\bf 1},  5793
  (1989).

\bibitem{hesd2}
A.~M. Cazabat, N. Fraysse, and F. Heslot, Colloids ans Surfaces {\bf 52},  1
  (1991).

\bibitem{CazabatdeGennes}
P. de~Gennes and A. Cazabat, C. R. Acad. Sci. {\bf 310},  1601  (1990).

\bibitem{Abragam}
D.~B. Abragam, P. Collet, J.~D. Coninck, and F. Dunlop, Phys. Rev. Lett. {\bf
  65},  195  (1990).

\bibitem{DS}
S. Burlatsky, G. Oshanin, A. Mogutov, and M. Moreau, Phys. Lett. A {\bf 166},
  230   (1992).

\bibitem{BOMR1}
S. Burlatsky, G. Oshanin, M. Moreau, and W. Reinhardt, to be subm.  (1995).

\bibitem{BOMR}
S.F. Burlatsky, W.P. Reinhardt, 
G.S. Oshanin, and M. Moreau 
{\it Bulletin of APS} {\bf 40} 301
 (1995)

\bibitem{BOMCR}
S. Burlatsky, A.M. Cazabat, G.S. Oshanin, M. Moreau and W.P. Reinhardt, To be publ.  (1995).

\bibitem{note}
Calculation of the numerical prefactors in some cases requares more elaborated
  theory. 


\end{thebibliography}

\begin{figure}
\caption{Sketch of experimental set-up. } \label{eseutup} 
\caption{Mean squared - diamonds, and squared mean - triangles,
diplacement of a single ball vs time,
averaged over $30$ trials.}
\label{tsetup}
\caption{The experimental dependence of mean number of balls on square root
of
time.
Triangles - non magnetic balls, $n=8$; squares - non magnetic balls, $n=4$;
diamonds - magnetic balls, solid lines - linear regression.} \label
{ballonsqrt}
\caption{Mean displacement of the rightmost ball vs square root of time.
Triangles - experimental data for  non magnetic balls, $n=8$; squares -
experimental data for non magnetic balls, $n=4$; diamonds -
experimental data for magnetic balls, solid lines - linear regression.}
\label{displsqrt}
\caption{Time vs number of
balls which emerged from vertical stack. Squares
experimental
data for non magnetic balls. Solid line - theoretical dependence predicted
by
Eq.~(8)
for $t<t_0$, dashed line - theoretical curve, predicted by Eq.~(12). }
\label
{extter}
\caption{Mean square displacement - squares, and square of mean displacement
-
triangles,
vs time for $n=8$ and non magnetic balls.} \label{msqsqm}
\caption{Numerical data: mean number of balls on the horizontal rack,
$p$ grows from 0.5 to 0.8.} \label{simnpart}
\caption{Numerical data: mean displacement averaged for all balls on the
horizontal stack,
$p$ grows from 0.5 to 0.8. } \label{simdispl}
\caption{Numerical data: mean displacement of the rightmost ball along the
horizontal
stack,
$p$ grows from 0.5 to 0.8.}  \label{simmaxdispl}
\end{figure}

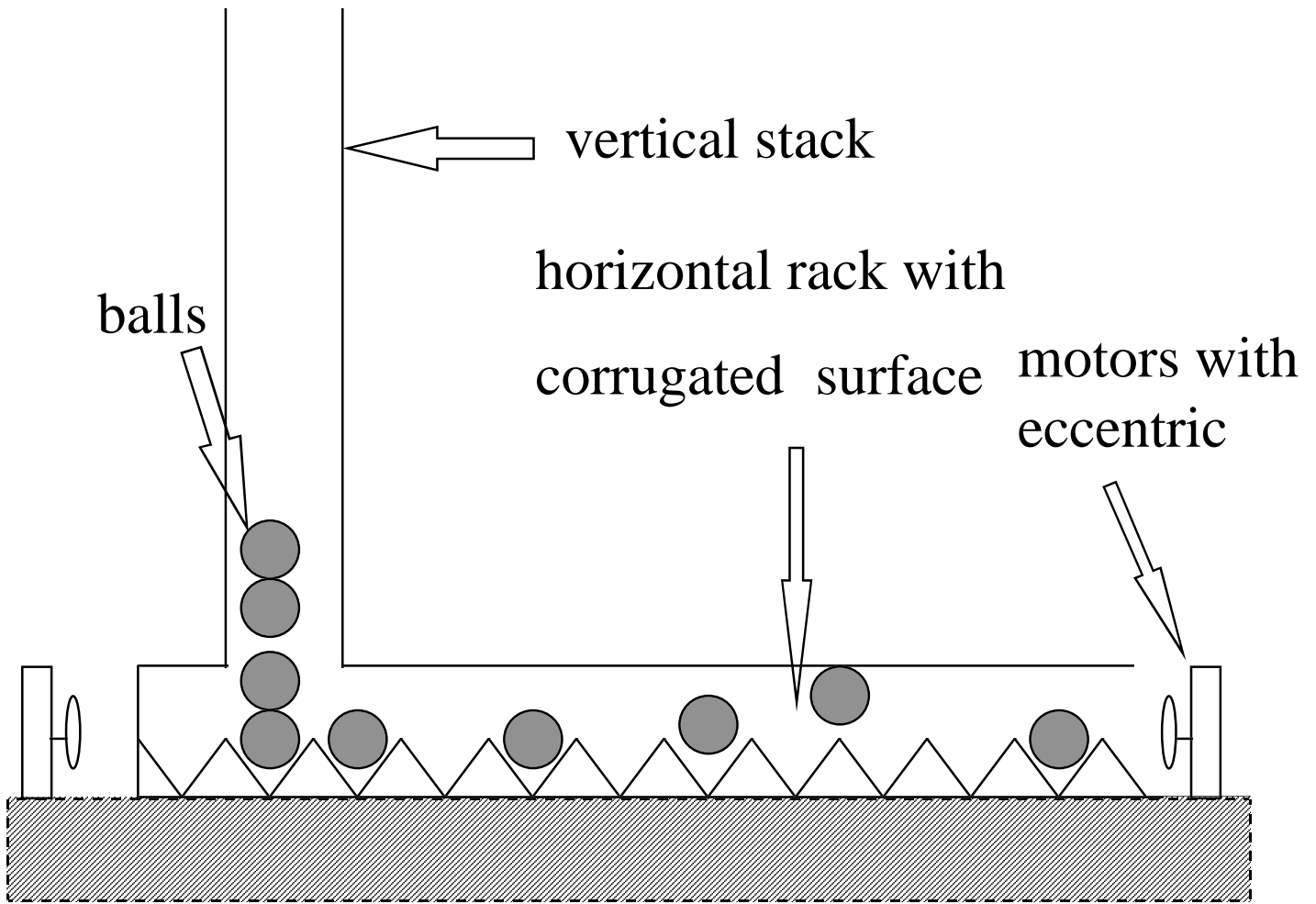
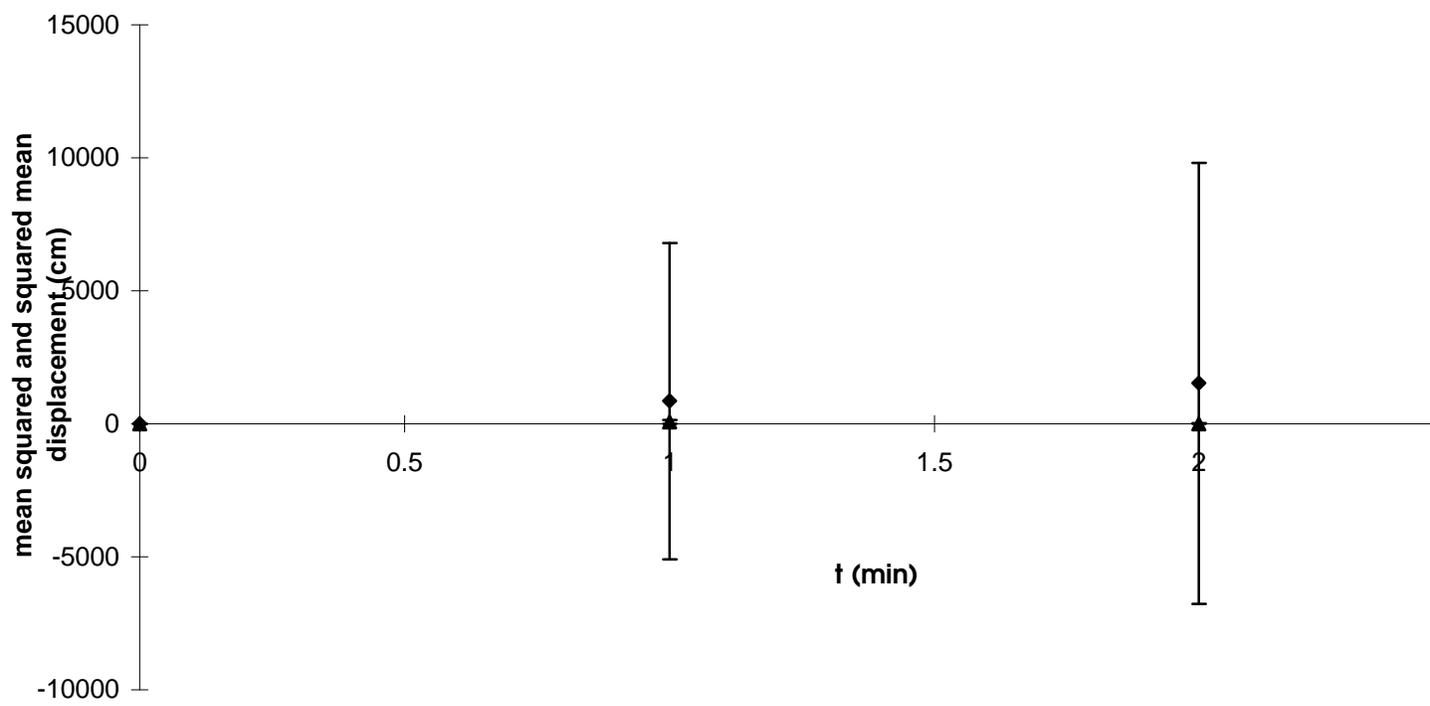
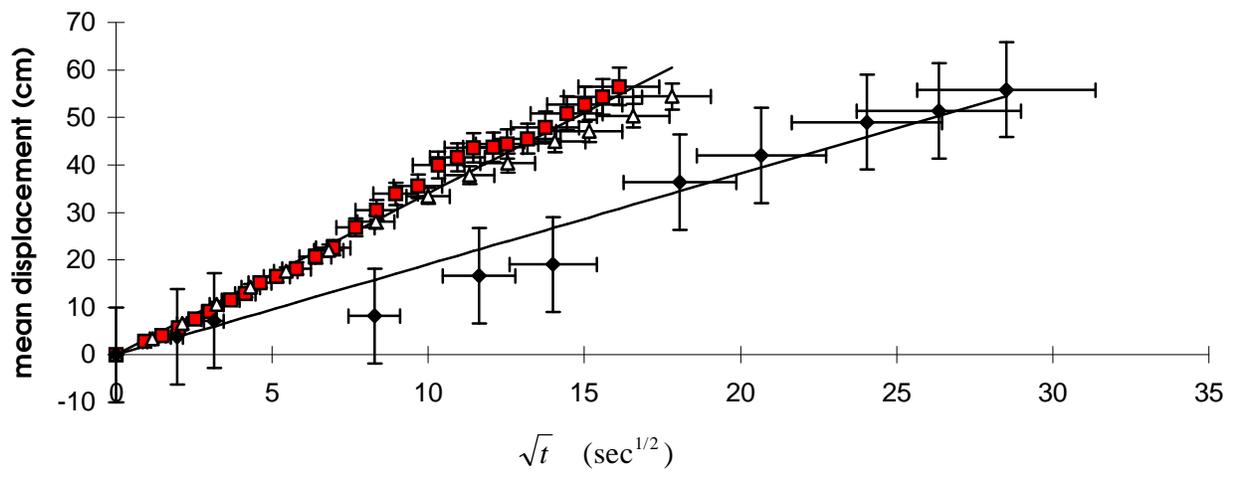
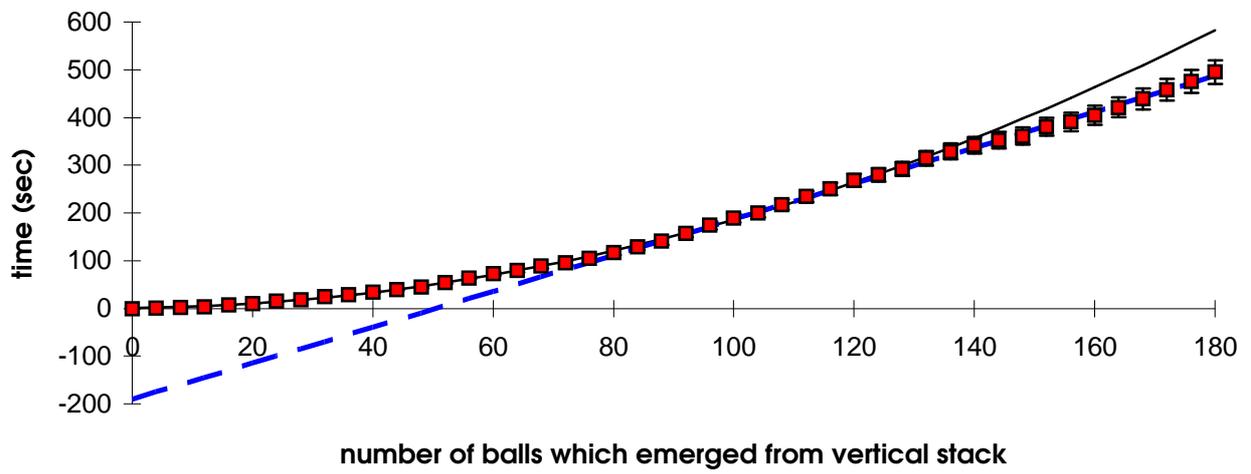
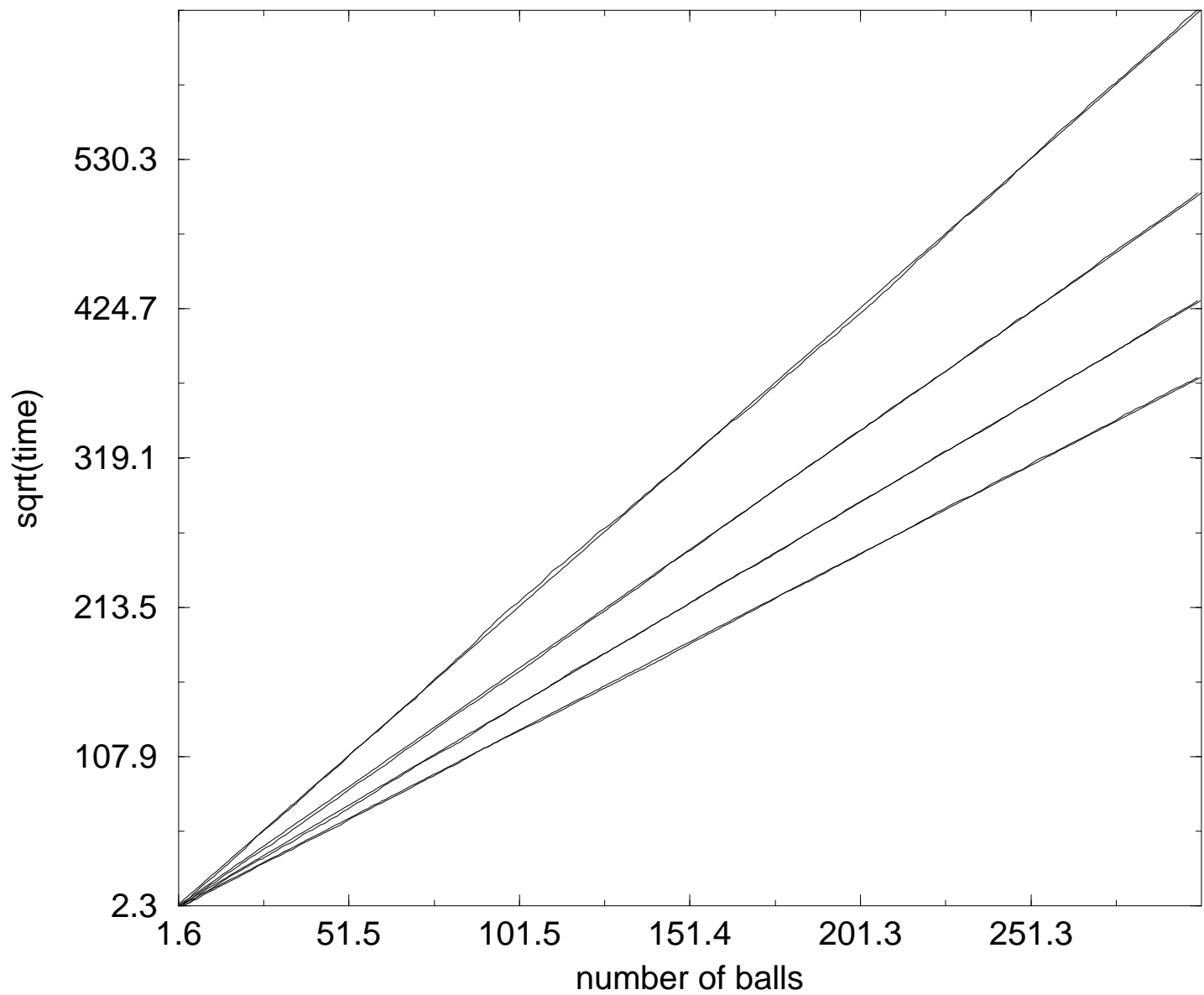
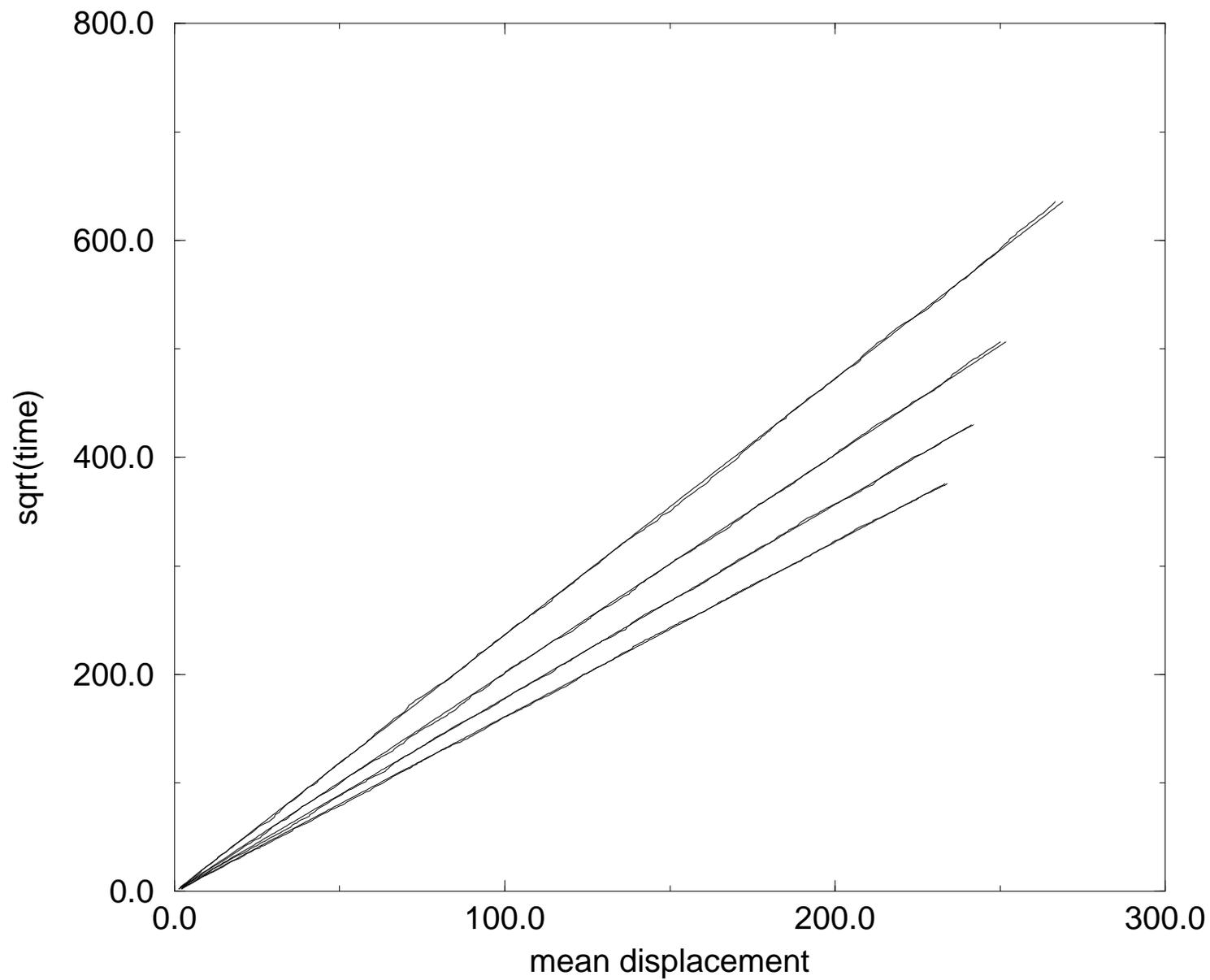
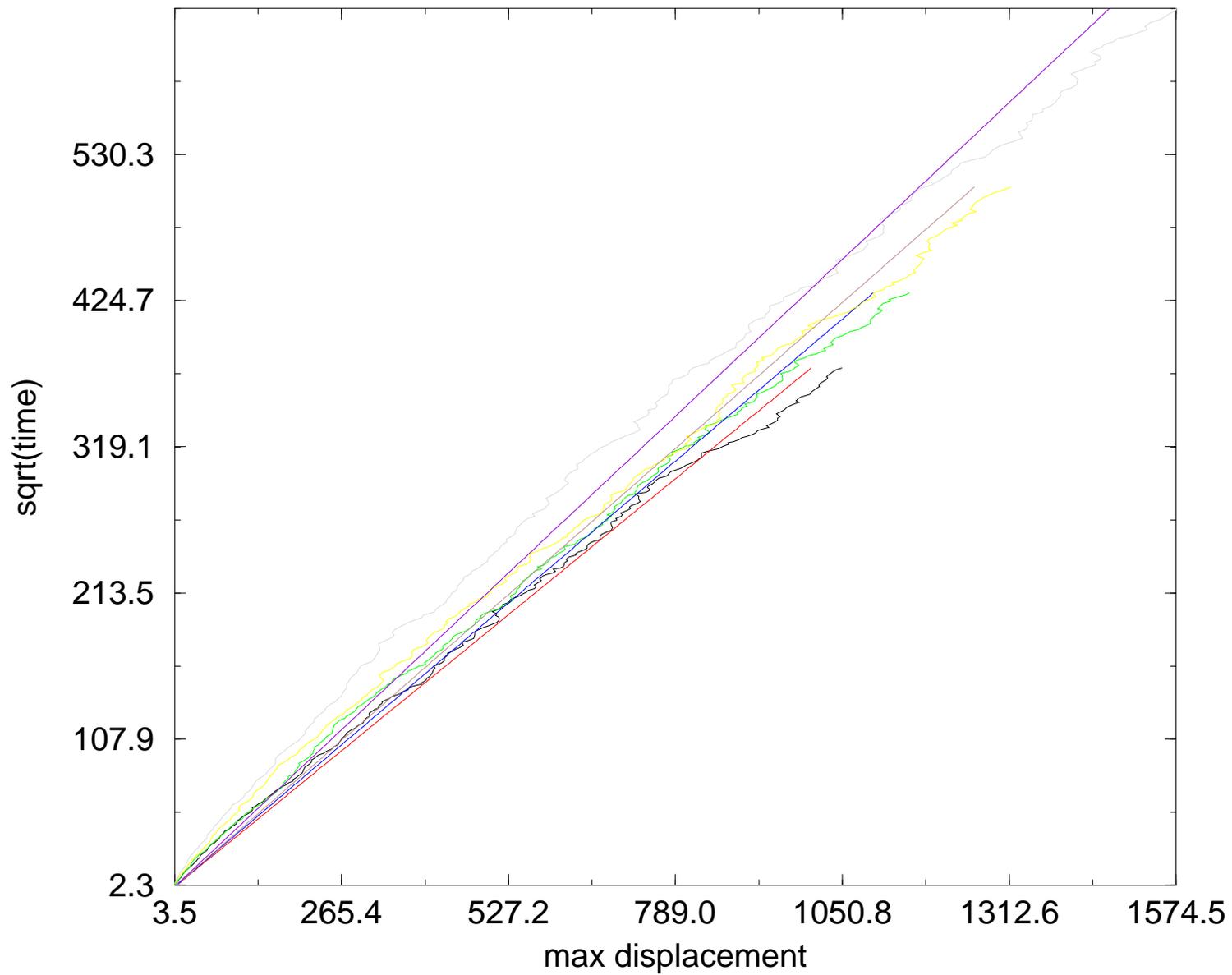
\end{document}